\documentclass[10pt]{article}

\usepackage{graphicx}

\begin{document}

\title{Bianchi type III models with anisotropic dark energy}
\date{} 
\author{\"{O}zg\"{u}r Akarsu\footnote{E-Mail: ozgur.akarsu@mail.ege.edu.tr} \and Can Battal K{\i}l{\i}n\c{c}\footnote{E-Mail: can.kilinc@ege.edu.tr}}
\maketitle
\begin{center}
\vskip-1cm
\textit{Ege University, Faculty of Science, Dept. of Astronomy and Space Sciences, 35100 Bornova, {\.I}zmir/Turkey.}
\end{center}

\begin{abstract}
The general form of the anisotropy parameter of the expansion for Bianchi type-III metric is obtained in the presence of a single diagonal imperfect fluid with a dynamically anisotropic equation of state parameter and a dynamical energy density in general relativity. A special law is assumed for the anisotropy of the fluid which reduces the anisotropy parameter of the expansion to a simple form ($\Delta\propto H^{-2}V^{-2}$, where $\Delta$ is the anisotropy parameter, $H$ is the mean Hubble parameter and $V$ is the volume of the universe). The exact solutions of the Einstein field equations, under the assumption on the anisotropy of the fluid, are obtained for exponential and power-law volumetric expansions. The isotropy of the fluid, space and expansion are examined. It is observed that the universe can approach to isotropy monotonically even in the presence of an anisotropic fluid. The anisotropy of the fluid also isotropizes at later times for accelerating models and evolves into the well-known cosmological constant in the model for exponential volumetric expansion.

\begin{flushleft}
\end{flushleft}

\end{abstract}
\section{Introduction}
\label{intro}

Some large-angle anomalies, which appear to indicate violation of the statistical isotropy, have been found in the cosmic microwave background (CMB) radiation \cite{{Smoot},{Bennett},{Hinshaw03},{Hinshaw07},{Hinshaw09}} (see \cite{{Eriksen},{Oliveira},{Cruz},{Hansen},{Land},{Copi},{Copi2}} for the anomalies). Non-trivial geometry, i.e., broken spherical symmetry, of the universe seems as the most promising explanation of these anomalies (see \cite{{Eriksen},{Copi},{Copi2}}). For instance, Jaffe et al. \cite{{Jaffe1},{Jaffe2},{Jaffe3}} showed that removing a Bianchi component from the WMAP initial data release can account for several large-angle anomalies and leave a statistically isotropic sky. The low value of the quadrupole moment, which has been known since the first COBE \cite{{Smoot},{Bennett}} results, is the most apparent of these anomalies and still exist in the high resolution WMAP data \cite{{Hinshaw03},{Hinshaw07},{Hinshaw09}}. Campanelli et al. \cite{{Campanelli06},{Campanelli07}} showed that, in Bianchi type-I framework, allowing the universe to be plane-symmetric with eccentricity (regardless of the origin) at decoupling of order $10^{-2}$ can resolve the quadrupole problem without effecting higher multipoles. They also concluded that WMAP data require an addition to the standart cosmological model that resembles the Bianchi morphology, which are homogeneous but not necessarily isotropic (see \cite{{Ellis1},{Ellis2}} for review on Bianchi metrics). Thus somehow in the cosmological models, the universe should have achieved a slightly anisotropic geometry inspite of the inflation. One may classify the models according to whether this occurs at an early time or at late times of the universe. In the context of the former class, the generic inflationary models can be modified in a way to end inflation with slightly anisotropic geometry, e.g. \cite{{Campanelli06},{Campanelli07}}. In the context of the latter class, the isotropy of the space that achieved during inflation can be distorted in the late time acceleration of the universe by modifiying the dark energy (DE), e.g. \cite{{Koivisto3},{Koivisto1},{Rodrigues}}.

In generic inflationary models \cite{{Guth81},{Sato},{Linde},{Albrecht}} characterized by an accelerated expansion in the early universe, space is assumed to be homegenous and isotropic from the begining and inflation is driven by a scalar field, which is isotropic. Some authors also studied scalar field models in Bianchi type space-times (e.g. \cite{{Feinstein93},{Aguirregabiria93}}). However, any possible anisotropy of the Bianchi metrics should have been died away during the inflation \cite{{Ellis1},{Ellis2}}. On the other hand, scalar fields may be replaced by vector fields, which give rise to anisotropic EoS parameter, once the metric is generalized to Bianchi type metrics. Inflation by using vector field first introduced by Ford \cite{Ford}, nevertheless it was suffering from fine tuning problem. Recently, Koivisto and Mota \cite {{Koivisto4}} have considered several new classes of viable vector field alternatives to the inflaton within in the Bianchi type-I framework. Golovnev et al. \cite{Golovnev} has constructed a succesful model, which either could give completely isotropic universe or slightly anisotropic universe at the end of inflation. Such models would also provide opportunity to construct more realistic inflationary models than the Bianchi type inflationary models which are driven by a scalar field and isotropize as evolved. Because in such models one generalizes not only the metric, but also the EoS parameter of the inflaton in accordance with the metric, and afterwards it can be showed whether the metric and/or the energy source evolves towards the isotropy.

On the other hand, these lowest multipoles represent the scale of the horizon at approximately the DE domination begins, since then it is natural to associate these anomalies with the present acceleration of the universe and the intrinsic nature of the DE rather than inflation (thus, inflaton). An anisotropic DE energy can derive an anisotropic late time acceleration and can break the isotropy that had been achieved during inflation. The paramount characteristic of the DE is a constant or slightly changing energy density as the universe expands, but we do not know the nature of the DE very well (see \cite{{Sahni1},{Sahni2},{Alam},{Sahni3},{Copeland},{Padmanabhan},{Turner},{Carroll}} for reviews on the DE). DE has conventionally been characterized by the equation of state (EoS) parameter $w=p/\rho$ which is not necessarily constant, where $\rho$ is the energy density and $p$ is the pressure. The simplest DE candidate is the vacuum energy ($w=-1$), which is mathematically equivalent to the cosmological constant ($\rm{\Lambda}$). The other conventional alternatives, which can be described by minimally coupled scalar fields, are quintessence ($w\geq-1$), phantom energy ($w\leq-1$) and quintom (that can cross from phantom region to quintessence region as evolved) and have time dependent EoS parameters. In all these models, DE is handled as an isotropic fluid. However there is no a priori reason to assume the DE is isotropic in nature. In principle, the EoS parameter of DE may be generalized by determining the EoS parameter separately on each spatial axis in a consistent way with the considered metric, since the energy density is a scalar quantity but the pressure is vectorial. Such DE candidates can also be studied in the context of vectorial fields and such candidates have been proposed by several authors (see \cite{{Koivisto3},{Armendariz},{Kiselev},{Zimdahl},{Novello},{Wei}}). Unlike Robertson-Walker (RW) metric Bianchi type metrics can admit a DE that wields an anisotropic EoS parameter according to their characteristics. The cosmological data -from the large-scale structures \cite{Tegmark} and Type Ia supernovae \cite{{Riess},{Astier}} observations- do not rule out the possibility of an anisotropic DE either \cite{{Koivisto3},{Mota}}.

In last years cosmological models in the presence of an anisotropic DE within the Bianchi type-I framework have been studied by several authors. Rodrigues \cite{{Rodrigues}} has proposed a $\rm{\Lambda}$CDM cosmological model extension whose DE component preserves its nondynamical character but wields anisotropic vacuum pressure. Koivisto and Mota \cite {{Koivisto1}} have presented a two-fluid model in the presence of an anisotropic DE and perfect fluid which are interacting, and presented a vector field action for DE as an example of the possibility of an anisotropic DE. They have shown that such models are cosmologically viable and can explain the large-angle anomalies in the CMB. Koivisto and Mota \cite {{Koivisto2}} have investigated cosmologies where the accelerated expansion of the universe is driven by a field with an anisotropic equation of state by introducing two skewness parameters to quantify the deviation of pressure from isotropy. They have studied the dynamics of the background expansion and analyzed a special case of an anisotropic cosmological constant in detail. Akarsu and Kilinc \cite{Akarsu} have proposed a two-fluid model in the presence of a perfect fluid and dynamical DE which wields dynamical and anisotropic EoS parameter.

All of the above studies are based on the idea that an anisotropic fluid gives rise to an anisotropy in the expansion in Bianchi type-I space-time. However, an anisotropic fluid must not necessarily promote the anisotropy in the expansion. Candidates of such energy sources may also act so as to support isotropization of the expansion as has mentioned by Akarsu and Kilinc \cite{Akarsu} in relatively earlier times and as has shown in this study, within the Bianchi type-III framework, in the entire history of the universe. Thus, even if we observe an isotropic expansion in the present universe we still cannot rule out possibility of DE with an anisotropic EoS.

Bianchi type-III cosmological models in the presence of DE have been studied in general relativity in the last thirty years. Moussiaux et al. \cite{Moussiaux} have given an exact particular solution of the Einstein field equations for vacuum with a cosmological constant. Lorenz \cite{Lorenz1} has presented a model with dust and a cosmological constant. Chakraborty and Chakraborty have given a bulk viscous cosmological model with variable gravitational constant ($G$) and $\rm{\Lambda}$ in \cite{Chakraborty}. Singh et al. \cite{SinghIII} have investigated a model with variable $G$ and $\rm{\Lambda}$ in the presence of perfect fluid by assuming a conservation law for the energy-momentum tensor. Recently, Tiwari \cite{Tiwari} has studied a model in the presence of perfect fluid and a time dependent $\rm{\Lambda}$ with constant deceleration parameter. Bali \& Tinker \cite{Bali} have investigated a model in the presence of bulk viscous barotropic fluid with variable $G$ and $\rm{\Lambda}$.

Letelier \cite{Letelier} has examined some two-fluid cosmological models, which have similar symmetries to those Bianchi type-III models, where the distinct four-velocity vectors of the two non-interacting perfect fluids generate an axially symmetric anisotropic pressure.

In this study we have first obtained the general form of the anisotropy parameter of the expansion for Bianchi type-III metric in the presence of a single diagonal imperfect fluid with a dynamically anisotropic EoS parameter and a dynamical energy density in general relativity. Then we have made an assumption on the anisotropy of the fluid in a way to reduce the anisotropy parameter of the expansion to a simple form and obtained a hypothetical fluid that obeys to a special form of an anisotropic EoS parameter. The exact solutions of the Einstein field equations have been obtained by assuming two different volumetric expansion laws in a way to cover all possible expansions: namely, exponential expansion and power-law expansion. Some features of the evolution of the metric and the dynamics of the anisotropic DE fluid have been examined. It has been shown that, in the Bianchi type-III framework, there can be solutions in which anisotropic fluid does not promote anisotropic expansion.

\section{Field equations}
\label{sec:1}
We consider the homegenous and anisotropic space-time described by Bianchi type-III metric in the form
\begin{equation}
ds^{2}=dt^{2}-A(t)^{2}dx^{2}-B(t)^{2}e^{-2\alpha x}dy^{2}-C(t)^{2}dz^{2}
\end{equation}
where $A(t)$, $B(t)$ and $C(t)$ are the scale factors (metric tensors) and functions of the cosmic time $t$, and $\alpha\neq0$ is a constant. (Bianchi type-I metric can be recovered by choosing $\alpha=0$, but the underlying Lie algebra of the isometry group of the Bianchi type-I and type-III metrics are completely different \cite{Ellis2}. This metric does not cover Robertson-Walker metric, but gets its closest form to RW metric when $A(t)=B(t)=C(t)$, thus we may talk about its approaching to isotropy, but not a total isotropization of this metric.)

The simplest generalization of the EoS parameter of a perfect fluid may be to determine the EoS parameter separately on each spatial axis by preserving the diagonal form of the energy-momentum tensor in a consistent way with the considered metric as disccused in the introduction. (In fact, within the Bianchi type-III framework we would allow the off-diagonal terms with ${T}_{1}^{2}={T}_{2}^{1}$ to be non-null. However, in this study, we are dealing only with an anisotropic fluid whose energy-momentum tensor is in diagonal form.) Thus, we may write down the energy-momentum tensor of the fluid in the following form,
\begin{equation}
{{T}_{\nu}}^{\mu}=\mathrm{diag}[{{T}_{0}}^{0},{{T}_{1}}^{1},{{T}_{2}}^{2},{{T}_{3}}^{3}].
\end{equation}
Then we may parametrize it as follows,
\begin{eqnarray}
{{T}_{\nu}}^{\mu}=\mathrm{diag}[\rho,-{p_{x}},-{p_{y}},-{p_{z}}]=\mathrm{diag}[1,-{w_{x}},-{w_{y}},-{w_{z}}]\rho\\
\nonumber =\mathrm{diag}[1,-w,-(w+\gamma),-(w+\delta)]\rho,
\end{eqnarray}
where $\rho$ is the energy density of the fluid; ${p_{x}}$, ${p_{y}}$ and ${p_{z}}$ are the pressures and  ${w_{x}}$, ${w_{y}}$ and ${w_{z}}$ are the directional EoS parameters on the $x$, $y$ and $z$ axes respectively; $w$ is the deviation-free EoS parameter of the fluid. We have parametrized the deviation from isotropy by setting ${w_{x}}=w$ and then introducing skewness parameters $\delta$ and $\gamma$ that are the deviations from $w$ respectively on the $y$ and $z$ axes. $w$, $\delta$ and $\gamma$ are not necesarilly constants and can be functions of the cosmic time $t$.

The Einstein field equations, in natural units ($8\pi G=1$ and $c=1$), are
\begin{equation}
G_{\mu\nu}=R_{\mu\nu}-\frac{1}{2}Rg_{\mu\nu}=-{T}_{\mu\nu}
\end{equation}
where $g_{\mu\nu}u^{\mu}u^{\nu}=1$; $u^{\mu}=(1,0,0,0)$ is the four-velocity vector; $R_{\mu\nu}$ is the Ricci tensor; $R$ is the Ricci scalar, ${T}_{\mu\nu}$ is the energy-momentum tensor. 

In a comoving coordinate system, Einstein's field equations (4), for the anisotropic Bianchi-III space-time (1), in case of (3), read as 
\begin{equation}
\frac{\dot{A}}{A}\frac{\dot{B}}{B}+\frac{\dot{A}}{A}\frac{\dot{C}}{C}+\frac{\dot{B}}{B}\frac{\dot{C}}{C}-\frac{\alpha^2}{A^{2}}=\rho,
\end{equation}
\begin{equation}
\frac{\ddot{B}}{B}+\frac{\ddot{C}}{C}+\frac{\dot{B}}{B}\frac{\dot{C}}{C}=-w\rho,
\end{equation}
\begin{equation}
\frac{\ddot{A}}{A}+\frac{\ddot{C}}{C}+\frac{\dot{A}}{A}\frac{\dot{C}}{C}=-(w+\delta)\rho,
\end{equation}
\begin{equation}
\frac{\ddot{A}}{A}+\frac{\ddot{B}}{B}+\frac{\dot{A}}{A}\frac{\dot{B}}{B}-\frac{\alpha^2}{A^{2}}=-(w+\gamma)\rho,
\end{equation}
\begin{equation}
\alpha\left(\frac{\dot{A}}{A}-\frac{\dot{B}}{B}\right)=0
\end{equation}
where the over dot denotes derivation with respect to the cosmic time $t$.

\section{General discussion on isotropization and the solution}
\label{sec:2}
The anisotropy of the expansion can be parametrized after defining the directional Hubble parameters and the mean Hubble parameter of the expansion. The directional Hubble parameters in the directions of $x$, $y$ and $z$ for the Bianchi type-III metric defined in (1) may be defined as follows,
\begin{equation}
H_{x}\equiv\frac{\dot{A}}{A}\textnormal{,}\qquad H_{y}\equiv\frac{\dot{B}}{B}\qquad \textnormal{and}\qquad H_{z}\equiv\frac{\dot{C}}{C}
\end{equation}
and the mean Hubble parameter is given as
\begin{equation}
H=\frac{1}{3}\frac{\dot{V}}{V}=\frac{1}{3}\left(\frac{\dot{A}}{A}+\frac{\dot{B}}{B}+\frac{\dot{C}}{C}\right),
\end{equation}
where $V=ABC$ is the volume of the universe. The anisotropy parameter of the expansion is defined as
\begin{equation}
\Delta\equiv\frac{1}{3}\sum_{i=1}^{3}\left(\frac{H_{i}-H}{H}\right)^{2},
\end{equation}
where $H_{i}$ (i=1,2,3) represent the directional Hubble parameters in the directions of $x$, $y$ and $z$ respectively.

$\Delta=0$ corresponds to isotropic expansion. The space approaches isotropy, in case of diagonal energy-momentum tensor (${T}^{0i}=0$, where $i=1,\,2,\,3$) if $\Delta\rightarrow 0$, $V\rightarrow +\infty$ and ${T}^{00}>0$ ($\rho>0$) as $t\rightarrow+\infty$ (see \cite{Collins} for details).

After giving the above definition (12), if we use only the Einstein field equations (5-9) without introducing any constraint, we can obtain the most general form of the anisotropy parameter of the expansion for Bianchi type-III model in the presence of an anisotropic fluid with a diagonal energy-momentum tensor in general relativity.

Solution of the Eq. (9) gives
\begin{equation}
B=c_{1}A,
\end{equation}
where $c_{1}$ is the positive constant of integration. We substitute (13) into (7), subtract the result from (6), and obtain that the skewness parameter on the $y$ axis is null,
\begin{equation}
\delta=0,
\end{equation}
which means that the directional EoS parameters, hence the pressures, on the $x$ and $y$ axes are equal. On the other hand, the deviation of the directional EoS parameter from $w$ on the $z$ axis, $\gamma$, is not constrained to be null by the Einstein field equations. Now, using (10), (11), and (13), anisotropy parameter of the expansion (12) can be reduced to
\begin{equation}
\Delta=\frac{2}{9}\frac{1}{H^{2}}\left(H_{x}-H_{z}\right)^{2}.
\end{equation}
$H_{x}-H_{z}$, the difference between the expansion rates on $x$ and $z$ axes, can be obtained by using the field equations.

The field equations are reduced to the equations below, when Eqs. (13) and (14) are substituted into (5-9),
\begin{equation}
\frac{{\dot{A}}^{2}}{A^{2}}+2\frac{\dot{A}}{A}\frac{\dot{C}}{C}-\frac{\alpha^2}{A^{2}}=\rho,
\end{equation}
\begin{equation}
\frac{\ddot{A}}{A}+\frac{\ddot{C}}{C}+\frac{\dot{A}}{A}\frac{\dot{C}}{C}=-w\rho,
\end{equation}
\begin{equation}
2\frac{\ddot{A}}{A}+{\frac{\dot{A}}{A}}^{2}-\frac{\alpha^2}{A^{2}}=-(w+\gamma)\rho.
\end{equation}
On solving the equation which is obtained by subtracting (17) from (18) we obtain
\begin{equation}
H_{x}-H_{z}=\frac{\dot{A}}{A}-\frac{\dot{C}}{C}=\frac{\lambda}{V}+\frac{1}{V}\int\left(\frac{\alpha^{2}}{A^{2}}-\gamma\rho\right)Vdt,
\end{equation}
where $\lambda$ is the real constant of integration and the term with $\gamma$ is the term that arises due to the possible intrinsic anisotropy of the fluid. Finally using (19) in (15) we obtain the anisotropy parameter of the expansion,
\begin{equation}
\Delta=\frac{2}{9}\frac{1}{H^{2}}\left[\lambda+\int \left(\frac{\alpha^{2}}{A^{2}}-\gamma\rho\right)Vdt\right]^{2}V^{-2}.
\end{equation}

The anisotropy parameter of the expansion can be reduced to the equation  below (21) for a Bianchi type-III cosmological model in the presence of a perfect (thus isotropic) fluid by choosing $\gamma=0$,
\begin{equation}
\Delta=\frac{2}{9}\frac{1}{H^{2}}\left[\lambda+\alpha^{2}\int\frac{V}{A^{2}}dt\right]^{2}V^{-2}.
\end{equation}

The integral term in (20) vanishes for
\begin{equation}
\gamma=\frac{\alpha^{2}}{\rho A^{2}},
\end{equation}
which also leads to the following energy-momentum tensor
\begin{eqnarray}
{{T}_{\nu}}^{\mu}=\mathrm{diag}\left[1,-w,-w,-w-\frac{\alpha^{2}}{\rho A^{2}}\right]\rho,
\end{eqnarray}
and reduces the anisotropy parameter of the expansion to the following form
\begin{equation}
\Delta=\frac{2}{9}\frac{\lambda^{2}}{H^{2}}V^{-2}.
\end{equation}
One can check that this behaviour of the $\Delta$ (24) we obtained by using an \textit{anisotropic fluid} (23) in Bianchi type-III space-time is equivalent to the ones that can be obtained similarly for Bianchi type-I and Bianchi type-V space-times by using any \textit{isotropic fluid}. Then one would see that the results we obtain for $\Delta$ in the models given below are equivalent to the ones obtained in \cite{{Kumar}} for Bianchi type-I and in \cite{{SinghRam},{SinghBaghel}} for Bianchi type-V space-time models in case of \textit{isotropic fluid}.

The vanishing of the integral term also reduces the difference between the expansion rates on $x$ and $z$ to the following form,
\begin{equation}
H_{x}-H_{z}=\frac{\lambda}{ABC}.
\end{equation}

We can also obtain the most general form of the energy density in Bianchi type-III framework by using the first field equation (5) and the definition of the anisotropy parameter of the expansion (12),
\begin{equation}
\rho =3H^{2}\left(1-\frac{\Delta}{2}\right)-\frac{\alpha^{2}}{A^{2}}.
\end{equation}

Below we present the exact solutions of the model in the presence of an anisotropic fluid described by the energy-momentum tensor given in (23), i.e., we use (22) in the Einstein field equations (16-18); 
\begin{equation}
\frac{{\dot{A}}^{2}}{A^{2}}+2\frac{\dot{A}}{A}\frac{\dot{C}}{C}=\rho+\frac{\alpha^{2}}{A^{2}}=(1+\gamma)\rho,
\end{equation}
\begin{equation}
\frac{\ddot{A}}{A}+\frac{\ddot{C}}{C}+\frac{\dot{A}}{A}\frac{\dot{C}}{C}=-w\rho,
\end{equation}
\begin{equation}
2\frac{\ddot{A}}{A}+\frac{{\dot{A}}^2}{A^{2}}=-w\rho.
\end{equation}
Now we have three linearly independent equations (27-29) and four unknown functions ($A$, $C$, $w$ and $\rho$), thus an extra equation is needed to solve the system completely. To do that we have used two different volumetric expansion laws,
\begin{equation}
V=c_{2}e^{3kt}
\end{equation}
and
\begin{equation}
V=c_{2}t^{3m},
\end{equation}
where $c_{2}$, $k$ and $m$ are positive constants. In this way, all possible expansion histories, the exponential expansion (30) and the power-law expansion (31), have been covered. The models with the exponential expansion and power-law for $m>1$ exhibit accelerating volumetric expansion. On the other hand while model for $m=1$ exhibits volumetric expansion with constant velocity, the models for $m<1$ exhibit decelerating volumetric expansion. Thus, phenomenologically, the anisotropic fluid we dealed here can be considered in the context of DE in the models with exponential expansion and power-law expansion for $m>1$.

\section{Model for exponential expansion}
\label{sec:3}
After solving the field equations (27-29) for the exponential volumetric expansion (30) by considering (13) and (25), we obtain the scale factors as follows
\begin{equation}
A=\left(\frac{c_{2}}{c_{1}c_{3}}\right)^{\frac{1}{2}}e^{kt-\frac{1}{9}\frac{\lambda}{kc_{2}}e^{-3kt}},
\end{equation}
\begin{equation}
B=\left(\frac{c_{1}c_{2}}{c_{3}}\right)^{\frac{1}{2}}e^{kt-\frac{1}{9}\frac{\lambda}{kc_{2}}e^{-3kt}},
\end{equation}
\begin{equation}
C={c_{3}}e^{kt+\frac{2}{9}\frac{\lambda}{kc_{2}}e^{-3kt}},
\end{equation}
where ${c_{3}}$ is a positive constant of integration. The mean Hubble parameter is,
\begin{equation}
H=k
\end{equation}
and the directional Hubble parameters on the $x$, $y$ and $z$ axes are, respectively,
\begin{equation}
H_{x}=H_{y}=k+\frac{1}{3}\frac{\lambda}{c_{2}}e^{-3kt}\qquad\textnormal{and}\qquad H_{z}=k-\frac{2}{3}\frac{\lambda}{c_{2}}e^{-3kt}.
\end{equation}
Using the directional and mean Hubble parameters in (15) we obtain 
\begin{equation}
\Delta=\frac{2}{9}\frac{\lambda^2}{{c_{2}}^{2}k^{2}}{e^{-6kt}}.
\end{equation}
One can check that this behaviour of the $\Delta$ is equivalent to the ones obtained for exponential expansion in Bianchi type-I \cite{Kumar} and Bianchi type-V \cite{{SinghRam},{SinghBaghel}} cosmological models with isotropic fluid.

We can obtain the energy density of the fluid by using the scale factors in (27);
\begin{equation}
\rho =3k^{2}-\frac{1}{3}\frac{\lambda^2}{{c_{2}}^2}e^{-6kt}-\alpha^2\frac{{c_{1}}{c_{3}}}{c_{2}}e^{-2kt+\frac{2}{9}\frac{\lambda}{kc_{2}}e^{-3kt}}.
\end{equation}
The deviation-free part of the anisotropic EoS parameter may be obtained by using (32) and (38) in (29);
\begin{equation}
w=\frac{\lambda^2+9{{c_{2}}^2}k^2e^{6kt}}{\lambda^2-9{{c_{2}}^2}k^2e^{6kt}+3\alpha^2{c_{1}}{c_{2}}{c_{3}}e^{4kt+\frac{2}{9}\frac{\lambda}{kc_{2}}e^{-3kt}}}.
\end{equation}
While the $\delta$ is already found to be zero, we can obtain $\gamma$ by using (32) and (38) in (22) as follows
\begin{equation}
\gamma =-\frac{3\alpha^2{c_{1}}{c_{2}}{c_{3}} e^{4kt+\frac{2}{9}\frac{\lambda}{kc_{1}}e^{-3kt}}}{\lambda^2-9{{c_{2}}^2}k^2e^{6kt}+3\alpha^2{c_{1}}{c_{2}} {c_{3}}e^{4kt+\frac{2}{9}\frac{\lambda}{kc_{2}}e^{-3kt}}}.
\end{equation}

The anisotropy of the expansion ($\Delta$) is not promoted by the anisotropy of the fluid and decreases to null exponentially as $t$ increases. The space approaches to isotropy in this model, since $\Delta\rightarrow 0$, $V\rightarrow\infty$ and $\rho>0$ as $t\rightarrow\infty$.

Both terms with $\lambda$ and $\alpha$ contribute to the energy density of the fluid $\rho$ negatively. The energy density ($\rho$), the deviation-free EoS parameter ($w$) and the skewness parameter ($\gamma$) are dynamical. As $t\rightarrow\infty$, the anisotropic fluid isotropizes and mimics the vacuum energy, which is mathematically equivalent to the cosmological constant ($\rm{\Lambda}$), i.e., $\gamma\rightarrow 0$, $w\rightarrow -1$ and $\rho\rightarrow 3k^{2}$.

One can observe that the universe approaches to isotropy monotonically even in the presence of the anisotropic fluid, and the anisotropic fluid isotropizes and evolves to the cosmological constant in case of exponential volumetric expansion. These observations are worth to pay attention, since we are inclined to think that anisotropy in an energy source gives rise to increase anisotropy in the expansion.

\section{Model for power-law expansion}
\label{sec:4}
After solving the field equations (27-29) for the power-law volumetric expansion (31) by considering (13) and (25), we obtain the scale factors as follows
\begin{equation}
A=\left(\frac{c_{2}}{c_{1}c_{3}}\right)^{\frac{1}{2}}t{^m} e^{-\frac{1}{3}\frac{\lambda}{c_{1}}\frac{t^{1-3m}}{3m-1}},
\end{equation}
\begin{equation}
B=\left(\frac{c_{1}c_{2}}{c_{3}}\right)^{\frac{1}{2}}t^{m} e^{-\frac{1}{3}\frac{\lambda}{c_{1}}\frac{t^{1-3m}}{3m-1}},
\end{equation}
\begin{equation}
C={c_{3}}t^{m} e^{\frac{2}{3}\frac{\lambda}{c_{1}}\frac{t^{1-3m}}{3m-1}},
\end{equation}
where ${c_{3}}$ is a positive constant of integration. The mean Hubble parameter is
\begin{equation}
H=\frac{m}{t}
\end{equation}
and the directional Hubble parameters on the $x$, $y$ and $z$ axes are, respectively,
\begin{equation}
H_{x}=H_{y}=\frac{m}{t}+\frac{1}{3}\frac{\lambda}{c_{2}}t^{-3m}\qquad\textnormal{and}\qquad H_{z}=\frac{m}{t}-\frac{2}{3}\frac{\lambda}{c_{2}}t^{-3m}.
\end{equation}
Using the directional and mean Hubble parameters in (15) we obtain 
\begin{equation}
\Delta=\frac{2}{9}\frac{\lambda^2}{{c_{2}}^2}\frac{t^{2-6m}}{m^2}.
\end{equation}
One can check that this behaviour of the $\Delta$ is equivalent to the ones obtained for the models that correspond to the power-law expansion in Bianchi type-I \cite{Kumar} and Bianchi type-V \cite{{SinghRam},{SinghBaghel}} cosmological models with isotropic fluid.

We can obtain the energy density of the fluid by using the scale factors in (27);
\begin{equation}
\rho =3m^2t^{-2}-\frac{1}{3}\frac{\lambda^2}{{c_{2}}^2}t^{-6m}-\alpha^{2}\frac{c_{1}c_{3}}{c_{2}}t^{-2m}e^{\frac{2}{3}\frac{\lambda}{c_{2}}\frac{t^{1-3m}}{3m-1}}.
\end{equation}
The deviation-free part of the anisotropic EoS parameter may be obtained by using (41) and (47) in (29);
\begin{equation}
w=\frac{\lambda^{2}t^{2}+3m{c_{2}}^2(3m-2)t^{6m}}{\lambda^{2}t^{2}+3\alpha^{2}c_{1}c_{2}c_{3}t^{4m+2}e^{\frac{2}{3}\frac{\lambda}{c_{2}}\frac{t^{1-3m}}{3m-1}}-9m^{2}{c_{2}}^{2}t^{6m}}.
\end{equation}
While the $\delta$ is already found to be zero, we obtain $\gamma$ by using (41) and (47) in (22) as follows
\begin{equation}
\gamma=-\frac{3\alpha^{2}c_{1}c_{2}c_{3}t^{4m+2}e^{\frac{2}{3}\frac{\lambda}{c_{2}}\frac{t^{1-3m}}{3m-1}}}{\lambda^{2}t^{2}+3\alpha^{2}c_{1}c_{2}c_{3}t^{4m+2} e^{\frac{2}{3}\frac{\lambda}{c_{2}}\frac{t^{1-3m}}{3m-1}}-9m^{2}{c_{2}}^{2}t^{6m}}.
\end{equation}

The volume of the universe expands indefinitely for all values of $m$.

Anisotropy of the expansion ($\Delta$) is not promoted by the anisotropy of the fluid. It behaves monotonically, decays to zero for $m>1/3$ and diverges for $m<1/3$ as $t\rightarrow\infty$, and is constant for $m=1/3$. 

One can see that the terms with $\lambda$ and $\alpha$ contribute the energy density of the fluid ($\rho$) negatively, and thus we can determine which values of $m$ are convenient for which times of the universe by applying the condition $\rho>0$. If $\lambda$ is null, according to the condition on $\rho$, the models for $m<1$ may represent the relatively earlier times of the universe, the models for $m>1$ may represent the relatively later times of the universe and the model for $m=1$ may represent the entire universe provided that $3c_{2}>{\alpha}^{2}c_{1}c_{3}$. The cases for non-zero values of $\lambda$ should be examined seperately. The models for $0<m\leq1/3$ may represent the relatively earlier times of the universe, $1/3<m<1$ may represent the intermediate times of the universe and $m>1$ may represent the relatively later times of the universe. Finally the model for $m=1$ may represent the relatively later times of the universe if $3c_{2}>{\alpha}^{2}c_{1}c_{3}$, otherwise the intermediate times.

Thus, we may examine the behaviours of $\Delta$, $w$ and $\gamma$ as $t\rightarrow\infty$ only for $m\geq 1$. For $m>1$, $\Delta\rightarrow 0$ and $V\rightarrow\infty$ as $t\rightarrow\infty$, thus universe approaches to isotropy. $w\rightarrow-1+\frac{2}{3m}$ and $\gamma\rightarrow0$ as $t\rightarrow\infty$, which means that the EoS parameter of the fluid isotropizes and approaches a value in quintessence region with regard the value of $m$ at the later times of the universe for accelerating models. For the model $m=1$, under the condition mentioned in the previous paragraph, $w\rightarrow -(3-{\alpha}^{2}c_{1}c_{3}/c_{2})^{-1}$ and $\gamma\rightarrow ({\alpha}^{2}c_{1}c_{3})/(3c_{2}-{\alpha}^{2}c_{1}c_{3})$ as $t\rightarrow\infty$. Thus, the space approaches to isotropy since $\Delta\rightarrow 0$ and $V\rightarrow\infty$ as $t\rightarrow\infty$, but the fluid does not for the model $m=1$.

Similarly to the model with exponential expansion, the universe approaches to isotropy monotonically even in the presence of the anisotropic fluid for $m>1$ and for $m=1$ with appropriate values of the constants. However, the anisotropic fluid isotropizes only in the accelerating models ($m>1$) at later times of the universe and its EoS parameter evolves into the quintessence region. 

\section{Conclusion}
\label{conclusion}
We have first obtained the general form of the anisotropy parameter for the expansion of Bianchi type-III metric in the presence of a single imperfect fluid with a dynamically anisotropic equation of state (EoS) parameter and a dynamical energy density in general relativity. Then we have made an assumption on the anisotropy of the fluid in a way to reduce the anisotropy parameter of the expansion to a simple form and obtained a hypothetical fluid with an special anisotropic EoS parameter. The exact solutions of the Einstein field equations have been obtained by assuming two different volumetric expansion laws in a way to cover all possible expansions: namely, exponential expansion and power-law expansion. The anisotropy of the fluid, expansion and space have been examined.

It is observed that eventhough the fluid we used wields an anisotropic EoS parameter, its anisotropy does not promote anisotropy in the expansion. The expansion anisotropy decays to zero monotonically in the models with the exponential expansion and in the power-law expansion when $m>1/3$. The universe approaches to isotropy in the accelerating models (exponential expansion and the power-law models with $m>1$). The anisotropy of the fluid isotropizes at later times of the universe in the accelating models. The fluid evolves into the vacuum energy with $w=-1$, which is mathematically equivalent to the cosmological constant ($\rm{\Lambda}$) at the later times of the universe in the model for exponential expansion. The EoS parameter of the fluid evolves into the quintessence region at later times of the accelerating ($m>1$) universes in the power-law models. The anisotropic fluid we used here can be considered in the context of dark energy (DE), at least phenomenologically, in the accelerating models.

This model is of interest because it shows that even in the presence of an anisotropic fluid, the universe can approach to isotropy monotonically and also the anisotropy of the fluid can isotropize in the accelerating models. In short, an accelerated expansion period isotropizes both the expansion anisotropy and the anisotropy of the fluid in our study. Thus, even if we observe an isotropic expansion in the present universe we still cannot rule out possibility of DE with an anisotropic EoS. Additionaly we can conclude that an anisotropic DE does not necessarilly distort the symmetry of the space, and consequently even if it turns out that spherical symmetry of the universe that achieved during inflation has not distorted in the later times of the universe, we can not rule out the possibility of an anisotropic nature of the DE at least in Bianchi type-III framework.

\begin{center}
\textit{Acknowledgments}
\end{center}
\"{O}zg\"{u}r Akarsu was supported in part by The Scientific and Technological Research Council of Turkey (T\"{U}B{\.I}TAK).

%
%

\end{document}